\documentclass[conference,letterpaper]{IEEEtran}
\usepackage{balance}
\usepackage{cite}
\usepackage{float}
\usepackage{multicol,multirow}
\usepackage{makecell,booktabs}
\usepackage{hyperref}
\hypersetup{    
    colorlinks=true,
    linkcolor=black,
    anchorcolor=black,
    citecolor=black,
    filecolor=black,
    menucolor=black,
    runcolor=black,
    urlcolor=black,
}

\usepackage{subcaption}
\usepackage[utf8]{inputenc}
\usepackage{graphicx}
\usepackage{amsfonts}
\usepackage{amssymb}
\usepackage{amsmath,mathtools}
\usepackage{array}
\usepackage{algorithm}
\usepackage{algpseudocode}

\newcolumntype{P}[1]{>{\centering\arraybackslash}p{#1}}
\newcolumntype{L}[1]{>{\raggedright\arraybackslash}p{#1}}
\newcolumntype{R}[1]{>{\raggedleft\arraybackslash}p{#1}}
\usepackage{color}
\usepackage[labelsep=period]{caption}
\usepackage{bbm}
\usepackage{bm}
\usepackage{comment}
\usepackage{forest}
\usepackage{dblfloatfix}
\usepackage{etoolbox}
\usepackage{cancel}

\title{MPA-DNN: Projection-Aware Unsupervised Learning for Multi-period DC-OPF \\
\thanks{This work was supported by Basic Science Research Program through the National Research Foundation of Korea (NRF) funded by the Ministry of Education (No. RS-2023-00210018) and KENTECH Research Grant (202300008A).}}

\author{\IEEEauthorblockN{Yeomoon Kim}
\IEEEauthorblockA{\textit{Dept. of Energy Engineering} \\
\textit{Korea Institute of Energy Technology}\\
Naju, South Korea \\
moon217@kentech.ac.kr}
\and
\IEEEauthorblockN{Minsoo Kim}
\IEEEauthorblockA{\textit{Dept. of Energy Engineering} \\
\textit{Korea Institute of Energy Technology}\\
Naju, South Korea \\
minsookim@kentech.ac.kr}
\and
\IEEEauthorblockN{Jip Kim}
\IEEEauthorblockA{\textit{Dept. of Energy Engineering} \\
\textit{Korea Institute of Energy Technology}\\
Naju, South Korea \\
jipkim@kentech.ac.kr}}

\begin{document}
\IEEEoverridecommandlockouts
\maketitle
\IEEEpubidadjcol
\begin{abstract}
    Ensuring both feasibility and efficiency in optimal power flow (OPF) operations has become increasingly important in modern power systems with high penetrations of renewable energy and energy storage. While deep neural networks (DNNs) have emerged as promising fast surrogates for OPF solvers, they often fail to satisfy critical operational constraints, especially those involving inter-temporal coupling, such as generator ramping limits and energy storage operations. To deal with these issues, we propose a Multi-Period Projection-Aware Deep Neural Network (MPA-DNN) that incorporates a projection layer for multi-period dispatch into the network. By doing so, our model enforces physical feasibility through the projection, enabling end-to-end learning of constraint-compliant dispatch trajectories without relying on labeled data. Experimental results demonstrate that the proposed method achieves near-optimal performance while strictly satisfying all constraints in varying load conditions.\vspace{-2mm}
\end{abstract}

\section{Introduction}\label{Sec:Intro}
Solving the Optimal Power Flow (OPF) problem is a fundamental task in modern power system operations. It determines the most cost-effective dispatch of generation units while satisfying a range of constraints, including power balance, generator capacity, transmission limits. The increasing integration of renewable energy sources (RES) with high variability, energy storage systems (ESS), and large-scale interconnected networks has significantly increased the problem complexity \cite{igogo2021integrating}. Furthermore, multi-period OPF formulations that account for inter-temporal constraints, are computationally intensive.

To address this issue, DNN-based surrogate models have been investigated in recent studies to learn the mapping from demand profiles to generator schedules \cite{van2021machine}. These models offer orders-of-magnitude speedups compared to conventional optimization solvers, making them highly attractive for real-time decision-making in power system operation.

However, existing learning-based OPF methods face three major challenges. First, they often fail to guarantee feasibility with respect to physical and operational constraints. Second, most approaches rely on supervised learning, which requires costly ground-truth labels generated by conventional optimization solvers. Third, many methods neglect inter-temporal constraints which are essential for realistic multi-period scheduling.

To address the first challenge, several studies have incorporated the Karush-Kuhn-Tucker (KKT) conditions into the loss function to penalize constraint violations during training~\cite{nellikkath2021physics, fioretto2020predicting, zhang2021convex}. Such approaches quantify the performance of DNN-based OPF models by analyzing their worst-case violations of physical constraints and deviations from optimality. Although effective on training data, these methods often struggle to generalize to unseen test scenarios, particularly when operating conditions deviate from the training distribution.

Other approaches apply post-processing techniques to correct the neural network output after inference~\cite{huang2021deepopf}. However, such approaches provide no guarantees during training and may yield suboptimal or infeasible dispatch decisions in real-time operations. Recently, the projection-aware deep neural network (PA-DNN) was proposed to overcome this issue \cite{kim2022projection}, which enforces feasibility in DC-OPF via a embedded projection layer, such as OptNet \cite{amos2017optnet} and CVXPYLayer \cite{agrawal2019differentiable}. However, these works rely on supervised learning and are unable to overcome the second challenge. In other words, pre-solved solutions are required to train these models. Since generating ground-truth solutions across diverse operating conditions is particularly impractical for large-scale systems with high renewable penetration and ESS integration, repeatedly solving multi-period OPF under such scenarios is computationally prohibitive. Additionally, they are typically limited to static, single-period problems and fail to capture inter-temporal constraints such as generator ramping and ESS state-of-charge, leaving the third challenge unaddressed.

In this regard, we propose the Multi-Period Projection-Aware Deep Neural Network (MPA-DNN) to overcome these limitations. To deal with the first challenge, we integrate a differentiable projection layer that enforces hard feasibility during both training and inference. For the second challenge, we adopt an unsupervised loss function based on system-level cost, removing the dependence on supervised labels. Finally, we explicitly model inter-temporal constraints including generator ramping limits and ESS energy dynamic within the projection formulation to overcome the third challenge.

\begin{figure*}[t]
    \centering
    \includegraphics[width=0.8\textwidth]{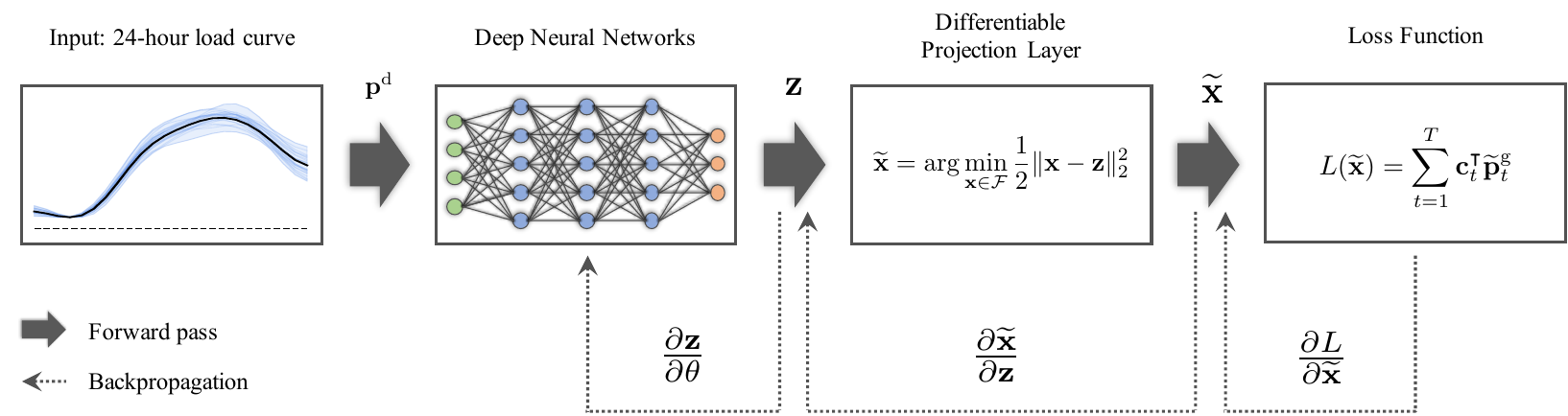}
    \caption{\small MPA-DNN architecture with a differentiable projection layer and gradient flow through the KKT system. The input is a 24-hour load profile vector, which is mapped to a raw generation vector $\mathbf{z}$ via a deep neural network. The differentiable projection layer solves a constrained QP to obtain a feasible dispatch $\widetilde{\mathbf{x}}$ by projecting $\mathbf{z}$ onto the feasible set $\mathcal{F}$ over the entire time horizon. The loss function is computed using $\widetilde{\mathbf{x}}$. The network parameters are updated through the projection layer by the backpropagation.}
    \label{fig:fig1}
    \vspace{-5mm}
\end{figure*}

We summarize our key contributions as follows:
\begin{itemize}
    \item We develop a fully unsupervised deep learning framework for multi-period DC-OPF, where generator schedules are optimized solely based on operational cost, without relying on labeled solutions from optimization solvers. This eliminates the burden of costly label generation while maintaining end-to-end training.
    \item We integrate a differentiable projection layer into the neural architecture that strictly enforces feasibility with respect to power balance, generator capacity, ramping constraints, and ESS dynamics over the entire time horizon. \color{black} This guarantees the network to produce physically feasible dispatch schedules\color{black}.
    \item We evaluate our model on the IEEE 39-bus test system with 24-hour load profiles and ESS integration. Experimental results show that our MPA-DNN achieves a near-optimal operational cost while strictly satisfying all constraints, and outperforms baseline models in both accuracy and feasibility under varying load conditions.
\end{itemize}

\section{Multi-Period DC Optimal Power Flow Model}
For a power system comprising $n_\mathrm{b}$ buses, $n_\mathrm{g}$ generators, $n_\mathrm{d}$ loads, $n_\mathrm{l}$ transmission lines, and $n_\mathrm{e}$ energy storage systems, the multi-period DC-OPF over a time horizon of $T$ periods is formulated as follows ($t \in \{1, \dots, T\}$):
\begin{subequations}\label{eq:multi-period_formulation}\begin{align}
    &\min_{\mathbf{p}_t^\mathrm{g}} && \sum_{t=1}^{T} \mathbf{c}_t^\intercal \mathbf{p}_t^\mathrm{g} \label{eq:sced_obj} \\
    &\text{s.t.} && \mathbf{1}^\intercal \mathbf{p}_t^\mathrm{g}  + \mathbf{1}^\intercal(\mathbf{p}_t^{\text{dis}} - \mathbf{p}_t^{\text{ch}}) = \mathbf{1}^\intercal \mathbf{p}_t^\mathrm{d}, \label{eq:sced_balance} \\
    & && \underline{\mathbf{p}}^\mathrm{g}  \leq \mathbf{p}_t^\mathrm{g}  \leq \overline{\mathbf{p}}^\mathrm{g} , \label{eq:sced_gen_limits} \\
    & && -\mathbf{R}^\text{d} \leq \mathbf{p}_t^\mathrm{g}  - \mathbf{p}_{t-1}^\mathrm{g}  \leq \mathbf{R}^\text{u}, \label{eq:sced_ramping} \\
    & && \underline{\mathbf{E}} \leq \mathbf{e}_t \leq \overline{\mathbf{E}}, \label{eq:sced_ess_bounds} \\
    & && \mathbf{e}_1 = 0.5\overline{\mathbf{E}}, \quad \mathbf{e}_T = 0.5\overline{\mathbf{E}}, \label{eq:sced_ess_initfinal} \\
    & && \mathbf{e}_t = \mathbf{e}_{t-1} + \mathbf{p}_t^{\text{ch}} \cdot \boldsymbol{\eta}^{\text{ch}} - \mathbf{p}_t^{\text{dis}} / \boldsymbol{\eta}^{\text{dis}}, \label{eq:sced_ess_energy} \\
    & && \mathbf{0} \leq \mathbf{p}_t^{\text{ch}} \leq \overline{\mathbf{p}}^{\text{ch}}, \quad \mathbf{0} \leq \mathbf{p}_t^{\text{dis}} \leq \overline{\mathbf{p}}^{\text{dis}}, \label{eq:sced_ess_power} \\
    & && \left| \boldsymbol{\Phi} \left( \mathbf{M}^\mathrm{g} \mathbf{p}_t^\mathrm{g} - \mathbf{M}^\mathrm{d} \mathbf{p}_t^\mathrm{d} + \mathbf{M}^{\mathrm{e}} ( \mathbf{p}_t^{\text{dis}} - \mathbf{p}_t^{\text{ch}} ) \right) \right| \leq \overline{\mathbf{p}}^\mathrm{l}, \label{eq:sced_lineflow}
\end{align}
\end{subequations}
where $\mathbf{c}_t \in \mathbb{R}^{n_\mathrm{g}}$ is the generation cost coefficient vector, $\mathbf{p}_t^\mathrm{g} \in \mathbb{R}^{n_\mathrm{g}}$ is the active power generation vector at time $t$, and $\mathbf{p}_t^\mathrm{d} \in \mathbb{R}^{n_\mathrm{d}}$ is the demand vector at time $t$. The ESS charging and discharging power vectors are given by $\mathbf{p}_t^{\text{ch}} \in \mathbb{R}^{n_\mathrm{e}}$ and $\mathbf{p}_t^{\text{dis}} \in \mathbb{R}^{n_\mathrm{e}}$, with corresponding efficiency vectors $\boldsymbol{\eta}^{\text{ch}} \in \mathbb{R}^{n_\mathrm{e}}$ and $\boldsymbol{\eta}^{\text{dis}} \in \mathbb{R}^{n_\mathrm{e}}$. The vectors $\underline{\mathbf{p}}^\mathrm{g} \in \mathbb{R}^{n_\mathrm{g}}$ and $\overline{\mathbf{p}}^\mathrm{g} \in \mathbb{R}^{n_\mathrm{g}}$ define the minimum and maximum generator limits, while $\mathbf{R}^\text{d} \in \mathbb{R}^{n_\mathrm{g}}$ and $\mathbf{R}^\text{u} \in \mathbb{R}^{n_\mathrm{g}}$ represent ramp-down and ramp-up limits. The SoC of ESS is denoted by $\mathbf{e}_t \in \mathbb{R}^{n_\mathrm{e}}$, constrained by $\underline{\mathbf{E}} \in \mathbb{R}^{n_\mathrm{e}}$ and $\overline{\mathbf{E}} \in \mathbb{R}^{n_\mathrm{e}}$. The generation shift factor (GSF) matrix $\boldsymbol{\Phi} \in \mathbb{R}^{n_\mathrm{l} \times n_\mathrm{b}}$ establishes a linear mapping from nodal power injections to line power flows. At each time step $t$, the matrices $\mathbf{M}^\mathrm{g} \in \mathbb{R}^{n_\mathrm{b} \times n_\mathrm{g}}$, $\mathbf{M}^\mathrm{d} \in \mathbb{R}^{n_\mathrm{b} \times n_\mathrm{d}}$, and $\mathbf{M}^{\mathrm{e}} \in \mathbb{R}^{n_\mathrm{b} \times n_\mathrm{e}}$ define the spatial allocation of generator outputs, demands, and ESS power exchange, respectively, to net nodal injections. The line capacity limits are represented by $\overline{\mathbf{p}}^\mathrm{l} \in \mathbb{R}^{n_\mathrm{l}}$.

\section{Methodology}\label{sec:Method}
\subsection{Overall Framework of MPA-DNN}
As illustrated in Fig.~\ref{fig:fig1}, the proposed framework consists of a DNN, a differentiable projection layer, and an unsupervised loss function. The DNN takes $\text{vec}(\mathbf{p}^\mathrm{d}) \in \mathbb{R}^{n_\mathrm{d}T}$ as input, where $\mathbf{p}^\mathrm{d} = [\mathbf{p}_1^\mathrm{d}, \cdots, \mathbf{p}_T^\mathrm{d}]\in\mathbb{R}^{n_\mathrm{d}\times T}$ and outputs $\text{vec}(\mathbf{z}) \in \mathbb{R}^{p T}$, where $\mathbf{z}\in\mathbb{R}^{p \times T}$ is unconstrained generation schedule, where $p = n_\mathrm{g} + 2n_\mathrm{e}$. Note that $\text{vec}(\cdot)$ is a column-wise stacking operator. For example, $\text{vec}(\mathbf{p}^\mathrm{d}) = [\mathbf{p}_1^{\mathrm{d}\intercal}, \cdots, \mathbf{p}_T^{\mathrm{d}\intercal}]^\intercal$.

To ensure physical and operational feasibility, $\mathbf{z}$ is projected onto the feasible set $\mathcal{F}$, which is defined by multi-period DC-OPF constraints as shown in Equation~\eqref{eq:multi-period_formulation}. 
Let $\mathbf{x}$ denote a stacked decision variable, which is defined as follows:
\begin{equation}
    \mathbf{x} =
    [\mathbf{p}^{\mathrm{g}\intercal},
    \mathbf{p}^{\text{ch}\intercal},
    \mathbf{p}^{\text{dis}\intercal}]^\intercal
    \in \mathbb{R}^{p \times T}, 
\end{equation}
where $\mathbf{p}^\mathrm{g}\in\mathbb{R}^{n_\mathrm{g}\times T}$, 
$\mathbf{p}^{\text{ch}}\in\mathbb{R}^{n_\mathrm{e}\times T}$, 
and $\mathbf{p}^{\text{dis}}\in\mathbb{R}^{n_\mathrm{e}\times T}$ denote the generator outputs, ESS charging power, and ESS discharging power over $T$ periods, respectively. 
Then, we solve the following optimization problem for the projection:
\begin{equation}
    \widetilde{\mathbf{x}} 
    = \arg\min_{\mathbf{x} \in \mathcal{F}} 
    \frac{1}{2}\| \mathbf{x} - \mathbf{z} \|_2^2,
    \label{eq:projection}
\end{equation}
where $\widetilde{\mathbf{x}}$ is a stacked optimal decision variable, which is defined as follows:
\begin{equation}
    \widetilde{\mathbf{x}} =
    [\widetilde{\mathbf{p}}^{\mathrm{g}\intercal},
    \widetilde{\mathbf{p}}^{\text{ch}\intercal},
    \widetilde{\mathbf{p}}^{\text{dis}\intercal}]^\intercal
    \in \mathbb{R}^{p \times T}, 
\end{equation}
Here, $\widetilde{\mathbf{p}}^\mathrm{g} \in \mathbb{R}^{n_\mathrm{g}\times T}$, 
$\widetilde{\mathbf{p}}^{\text{ch}}\in\mathbb{R}^{n_\mathrm{e}\times T}$, 
and $\widetilde{\mathbf{p}}^{\text{dis}}\in\mathbb{R}^{n_\mathrm{e}\times T}$ denote the projected generator outputs, obtained corresponding ESS charging power, and ESS discharging power over $T$ periods, respectively. \textcolor{black}{This optimization based projection layer plays a crucial role in enforcing constraints in every forward pass of the proposed MPA-DNN. By doing so, our proposed method guarantees that the model’s outputs remain feasible at all times for multi-period problems.}

After that, the model is trained in an unsupervised manner by minimizing the total generation cost as a loss function $L = \sum_{t=1}^T\mathbf{c}_t^\intercal \mathbf{\widetilde{p}}_t^{\mathrm{g}}$, which is evaluated from the projected output $\widetilde{\mathbf{p}}^\mathrm{g}$. \color{black} Note that the framework is equally applicable to quadratic or nonlinear cost formulations.\color{black}

\subsection{Backpropagation}
To train the model through the projection layer, we employ the chain rule to represent the backpropagation process. With the learnable parameters $\theta$ of the MPA-DNN, we have
\begin{equation}
    \frac{\partial L}{\partial \theta} = \frac{\partial L}{\partial \widetilde{\mathbf{x}}} \cdot \frac{\partial \widetilde{\mathbf{x}}}{\partial \mathbf{z}} \cdot \frac{\partial \mathbf{z}}{\partial \theta}. \label{eq:backpropagation}
\end{equation}
Since the loss function is defined as a linear function, the gradient with respect to the projected generation $\widetilde{\mathbf{x}}$ is computed analytically. In addition, $\partial \mathbf{z} / \partial \theta$ is determined from automatic differentiation, which is widely adopted in various deep learning frameworks \cite{paszke2019pytorch}. Then, $\partial \widetilde{\mathbf{x}} / \partial \mathbf{z}$ is obtained by differentiation of the projection layer. The simplified version of the projection (\ref{eq:projection}) is defined as:
\begin{subequations}\begin{align}
    \min_{\mathbf{x}} \quad & \tfrac{1}{2}\|\mathbf{x} - \mathbf{z}\|_2^2 \\
    \text{s.t.} \quad 
    & \mathbf{A} \, \text{vec}(\mathbf{x}) = \mathbf{b}, \\
    & \mathbf{G} \, \text{vec}(\mathbf{x}) \leq \mathbf{h},
\end{align}\label{eq:projection_simple}
\end{subequations}
\color{black}where $\mathbf{A}$ and $\mathbf{b}$ enforce power balance, and $\mathbf{G}$ and $\mathbf{h}$ represent operational limits. \color{black} Finally, by solving (\ref{eq:projection_simple}), the projection ensures that $\mathbf{p}^{\mathrm{g}}$ lies within the feasible region $\mathcal{F}$. \color{black}For readability, the detailed representations are provided in Appendix~A.\color{black}

\begin{algorithm}[t]
\caption{Training process of MPA-DNN}\label{alg:PADNN}
\begin{algorithmic}[1]
\State{\textbf{Inputs:}}
\Statex\begin{tabular}[t]{ll}
     \hspace{0.5em} $\mathcal{D}^\mathrm{d}$&$\begin{aligned}
     &\text{Set of $K$ active power demand samples}\\
     &\mathcal{D}^\mathrm{d} = \Big\{\mathbf{D}_{k,t}^\mathrm{d}:k\in\{1,...,K\}, t\in\{1,...,T\}\Big\}
     \end{aligned}$\\
     \hspace{0.5em} $M$& Maximum epoch\\
     \hspace{0.5em} $T$& Time horizon\\
     \hspace{0.5em} $\eta$ & Learning rate\\
     \hspace{0.5em} $L$ & Loss function
\end{tabular}
\State{Randomly initialize the NNs parameters $\mathbf{\theta}$}
\For{$m \in \{1,...,M\}$}
\For{$k\in\{1,...K\}$}
\For{$t \in \{1,...,T\}$}
\State{$\mathbf{p}^\mathrm{d} \gets \mathbf{D}_{k,t}^\mathrm{d}$}
\State{$\mathbf{z} \gets DNN(\mathbf{p}^\mathrm{d}, \mathbf{\theta})$}
\State{$\widetilde{\mathbf{x}} \gets \arg\min_{\mathbf{x} \in \mathcal{F}}\frac{1}{2}\|\mathbf{x} - \mathbf{z}\|_2^2$}
\vspace{1mm}
\State{$\frac{\partial L}{\partial\widetilde{\mathbf{x}}} \gets \text{calculate analytically from the loss function}$}
\State{$\frac{\partial\widetilde{\mathbf{x}}}{\partial\mathbf{z}} \gets \text{calculate via }
(\ref{eq:linear_sys2})$}
\State{$\mathbf{\theta} \gets \mathbf{\theta} - \eta\frac{\partial L}{\partial\widetilde{\mathbf{x}}}\frac{\partial\widetilde{\mathbf{x}}}{\partial\mathbf{z}}\frac{\partial\mathbf{z}}{\partial\mathbf{\theta}}$}
\EndFor
\EndFor
\EndFor
\State \Return{$\mathbf{\theta}$}
\end{algorithmic}
\end{algorithm}

Now, we formulate the Lagrange function from the given optimization problem (\ref{eq:projection_simple}) as follows:
\begin{equation}
    \mathcal{L}({\mathbf{x}}, \bm{\lambda}, \bm{\mu}) = \frac{1}{2}\|{\mathbf{x}} - \mathbf{z}\|_2^2 + \bm{\lambda}^\intercal(\mathbf{A}{\mathbf{x}} - \mathbf{b}) + \bm{\mu}^\intercal(\mathbf{G}{\mathbf{x}} - \mathbf{h}),
\end{equation}
where $\bm{\lambda}\in\mathbb{R}$ and $\bm{\mu}\in\mathbb{R}$ denote the dual variables corresponding to the equality and inequality constraints, respectively. The KKT conditions for this problem are given by:
\begin{subequations}
\begin{align}
    &\widetilde{\mathbf{x}} - \mathbf{z} + \mathbf{A}^\intercal \bm{\lambda}^* + \mathbf{G}^\intercal \bm{\mu}^* = 0, \\
    &\mathbf{A}\widetilde{\mathbf{x}} - \mathbf{b} = 0, \\
    &\mathrm{diag}(\bm{\mu}^*)(\mathbf{G}\widetilde{\mathbf{x}} - \mathbf{h}) = 0.
    \end{align}
\end{subequations}
By differentiating the KKT system with respect to $\mathbf{z}$, we obtain the sensitivity of the projection output to $\mathbf{z}$:
\begin{equation}
    \begin{bmatrix}
    \mathbf{I} & \mathbf{G}^\intercal & \mathbf{A}^\intercal \\
    \mathrm{diag}(\bm{\mu}^*)\mathbf{G} & \mathrm{diag}(\mathbf{G}\widetilde{\mathbf{x}} - \mathbf{h}) & \mathbf{0} \\
    \mathbf{A} & \mathbf{0} & \mathbf{0}
    \end{bmatrix}
    \begin{bmatrix}
    \partial \widetilde{\mathbf{x}} \\
    \partial \bm{\mu}^* \\
    \partial \bm{\lambda}^*
    \end{bmatrix}
    =
    \begin{bmatrix}
    \partial \mathbf{z} \\
    \mathbf{0} \\
    \mathbf{0}
    \end{bmatrix}. \label{eq:linear_sys1}
\end{equation}
Solving this linear system yields $\partial \widetilde{\mathbf{x}} / \partial \mathbf{z}$, which allows gradients to propagate through the projection layer:
\begin{equation}
    \begin{bmatrix}
    \partial \widetilde{\mathbf{x}} \\
    \partial \bm{\mu}^* \\
    \partial \bm{\lambda}^*
    \end{bmatrix}
    =
    \begin{bmatrix}
    \mathbf{I} & \mathbf{G}^\intercal & \mathbf{A}^\intercal \\
    \mathrm{diag}(\bm{\mu}^*)\mathbf{G} & \mathrm{diag}(\mathbf{G}\widetilde{\mathbf{x}} - \mathbf{h}) & \mathbf{0} \\
    \mathbf{A} & \mathbf{0} & \mathbf{0}
    \end{bmatrix}^{-1}
    \begin{bmatrix}
    \partial \mathbf{z} \\
    \mathbf{0} \\
    \mathbf{0}
    \end{bmatrix}. \label{eq:linear_sys2}
\end{equation}
Finally, the model parameters are updated according to:
\begin{equation}
    \theta’ = \theta - \eta \cdot \frac{\partial L}{\partial \widetilde{\mathbf{x}}} \cdot \frac{\partial \widetilde{\mathbf{x}}}{\partial \mathbf{z}} \cdot \frac{\partial \mathbf{z}}{\partial \theta}, \label{eq:update_theta}
\end{equation}
where $\eta$ denotes the learning rate. The overall training process of the proposed method is provided in Algorithm~\ref{alg:PADNN}.

\section{Case Study}
\subsection{Simulation Setup}
\subsubsection{Data Preparation}
The proposed model is evaluated on the IEEE 39-bus test system, which comprises 39 buses, 10 generators, and 46 transmission lines \cite{babaeinejadsarookolaee2019power}. To introduce inter-temporal coupling in the dispatch decisions, an ESS is installed at Bus~19. The ESS is configured with a charging/discharging power limit of 50~MW, an energy capacity of 200~MWh, and a round-trip efficiency of 90\%. The initial and final states of charge are fixed at 50\% of the total capacity.

Training data for the baselines with supervised learning is generated by solving DC-OPF problems using Gurobi. A total of 10{,}000 samples are generated and partitioned into training, validation, and test sets with a ratio of 5:3:2.

\subsubsection{Described Models}
We evaluate five baseline models summarized in Table~\ref{table:model_description}, with brief descriptions below:

\begin{itemize}
    \item \textbf{Solver (Gurobi}): The exact optimization-based benchmark solution, incorporating all physical and inter-temporal constraints.
    \item \textbf{SPA-DNN}\cite{kim2022projection}: Single-period PA-DNN trained in an unsupervised manner; does not account for inter-temporal constraints.
    \item \textbf{MPP-DNN(S)}\cite{pan2020deepopf}: Multi-period Post-Processing DNN trained with supervision; applies the projection step only at inference time to enforce feasibility.
    \item \textbf{MPA-DNN(S)}: Multi-period PA-DNN trained with supervised labels and end-to-end constraint enforcement via a differentiable projection layer.
    \item \textbf{MPA-DNN (Proposed)}: Our proposed model, a Multi-period PA-DNN trained without supervision. It enforces feasibility via a differentiable projection layer and minimizes system cost through an unsupervised loss.
\end{itemize}
\begin{table}[t]
    \centering
    \captionsetup{justification=centering, labelsep=period, font=footnotesize, textfont=sc}
    \caption{Comparison of baseline models.}
    \label{table:model_description}
    \begin{tabular}{l|c|c}
        \toprule
        \textbf{Model} & \textbf{\makecell{Inter-temporal \\ Constraints}} & \textbf{Supervision Type} \\
        \midrule\midrule
        Solver (Gurobi) & \checkmark & -- \\
        SPA-DNN & $\times$ & Unsupervised \\
        MPP-DNN(S) & \checkmark & Supervised (inference only) \\
        MPA-DNN(S) & \checkmark & Supervised \\
        MPA-DNN (Proposed) & \checkmark & Unsupervised \\
        \bottomrule
    \end{tabular}
    \vspace{-2mm}
\end{table}
For fair comparison, all learning-based models use the same DNN architecture, optimizer settings, and training dataset. Specifically, the network consists of \( h = 3 \) hidden layers, each with \( n_h = 40 \) neurons and exponential linear unit (ELU) activation function. The output layer size corresponds to the number of generators $n_\mathrm{g}$, while the input layer size is determined by the number of buses $n_\mathrm{b}$. All models are trained using the ADAM optimizer with a learning rate of \(10^{-4}\).

\subsection{Overall Performance Comparisons}
\begin{table}[t]
    \centering
    \captionsetup{justification=centering, labelsep=period, font=footnotesize, textfont=sc}
    \caption{Generalization performance across load scales. All the methods are trained at load scales of 1.000, and tested at 0\%, +2.5\%, and +5.0\%.}
    \label{table:overall_perform}
    \begin{tabular}{c|l|c|c}
        \toprule
        \textbf{\makecell{Load scale}} & \textbf{\makecell{Method}}&\textbf{\makecell{MAE (p.u.)}} & \textbf{\makecell{Opt. Gap (\%)}} \\
        \midrule\midrule
        \multirow{4}{*}{1.000} 
            & Solver (Gurobi) & 0.0000 & 0.0000 \\
            & MPP-DNN(S) & 0.0320 & 0.1333 \\
            & MPA-DNN(S) & 0.0175 & 0.0272 \\
            & \textbf{MPA-DNN (Proposed)} & \textbf{0.0160} & \textbf{0.0236} \\
        \midrule
        \multirow{4}{*}{1.025} 
            & Solver (Gurobi) & 0.0000 & 0.0000 \\
            & MPP-DNN(S) & 0.0337 & 0.1436 \\
            & MPA-DNN(S) & 0.0210 & 0.0242 \\
            & \textbf{MPA-DNN (Proposed)} & \textbf{0.0157} & \textbf{0.0209} \\
        \midrule
        \multirow{4}{*}{1.050} 
            & Solver (Gurobi) & 0.0000 & 0.0000 \\
            & MPP-DNN(S) & 0.0349 & 0.1725 \\
            & MPA-DNN(S) & 0.0296 & 0.0321 \\
            & \textbf{MPA-DNN (Proposed)} & \textbf{0.0155} & \textbf{0.0179} \\
        \bottomrule
    \end{tabular}
    \vspace{-5mm}
\end{table}

We summarize the mean absolute error (MAE) and the optimality gap of all methods at three load levels in Table \ref{table:overall_perform}. Each network is trained once on the nominal 1.000 scale and then evaluated, without retraining, on test cases whose demands are uniformly increased by +2.5 \% (1.025) and +5.0 \% (1.050). These scenarios provide evaluations of the models' ability to generalize to untrained operating conditions\color{black}, consistent with prior approaches to studying distribution shift in OPF learning \cite{venzke2020learning}.\color{black}

Across the three load scales, MPA-DNN records the lowest MAE ($\leq$ 0.016 p.u.) and keeps the optimality gap below 0.024\%, outperforming every baseline at both nominal and increased demand. In contrast, the MPP-DNN(S) model, which applies the projection layer only during inference, shows an increasing optimality gap under different load scales. These results are consistent with the findings of \cite{kim2022projection}, which shows that the post-processing model underperformed the projection-aware approach. Consistently, the supervised variant MPA-DNN(S) surpasses MPP-DNN(S), and the fully unsupervised MPA-DNN (proposed method) pushes performance even further by learning solely from the generation cost while enforcing the feasibility throughout the training.

\subsection{Necessity of Inter-temporal Constraints}
Table \ref{table:ramping_vio} shows ramping-constraint violations over 2,000 test samples. Each model is trained only once on the nominal 1.000 load dataset and then evaluated at different load levels, as done in Table~{\ref{table:overall_perform}. The proposed MPA-DNN satisfies every ramp limit across all scenarios, confirming that the embedded projection layer enforces ramping physics during training and remains effective even for the untrained test dataset with different load scales.
By contrast, SPA-DNN, which optimizes each hour independently and ignores ramp coupling, violates the limits more frequently as the load scale increases (e.g., 13 cases (0.65\%) at 1.000, 87 cases (4.35\%) at 1.025, and 140 cases (7.00\%) at 1.050).

\begin{table}[t]
    \centering
    \captionsetup{justification=centering, labelsep=period, font=footnotesize, textfont=sc}
    \caption{Ramping violations across load scales. All the methods are trained at load scales of 1.000, and tested at 0\%, +2.5\%, and +5.0\%.}
    \label{table:ramping_vio}
    \begin{tabular}{c|c|c}
        \toprule
        \textbf{Load Scale} & \textbf{SPA-DNN} & \textbf{MPA-DNN (Proposed)} \\
        \midrule\midrule
        1.000 & 13 / 2000 & \textbf{0 / 2000} \\
        1.025 & 87 / 2000 & \textbf{0 / 2000} \\
        1.050 & 140 / 2000 & \textbf{0 / 2000} \\
        \bottomrule
    \end{tabular}
    \vspace{-2mm}
\end{table}

These findings emphasize the necessity of incorporating inter-temporal constraints, such as ramping limits, into DNN models considering multi-period dispatch. By embedding a differentiable projection layer into the training process, the proposed MPA-DNN ensures feasibility throughout the entire prediction horizon. In contrast, the dispatch schedules generated by SPA-DNN are frequently infeasible and therefore have limited applicability in real-world power system operations.

\begin{figure}[t]
    \centering
    \includegraphics[width=0.85\columnwidth]{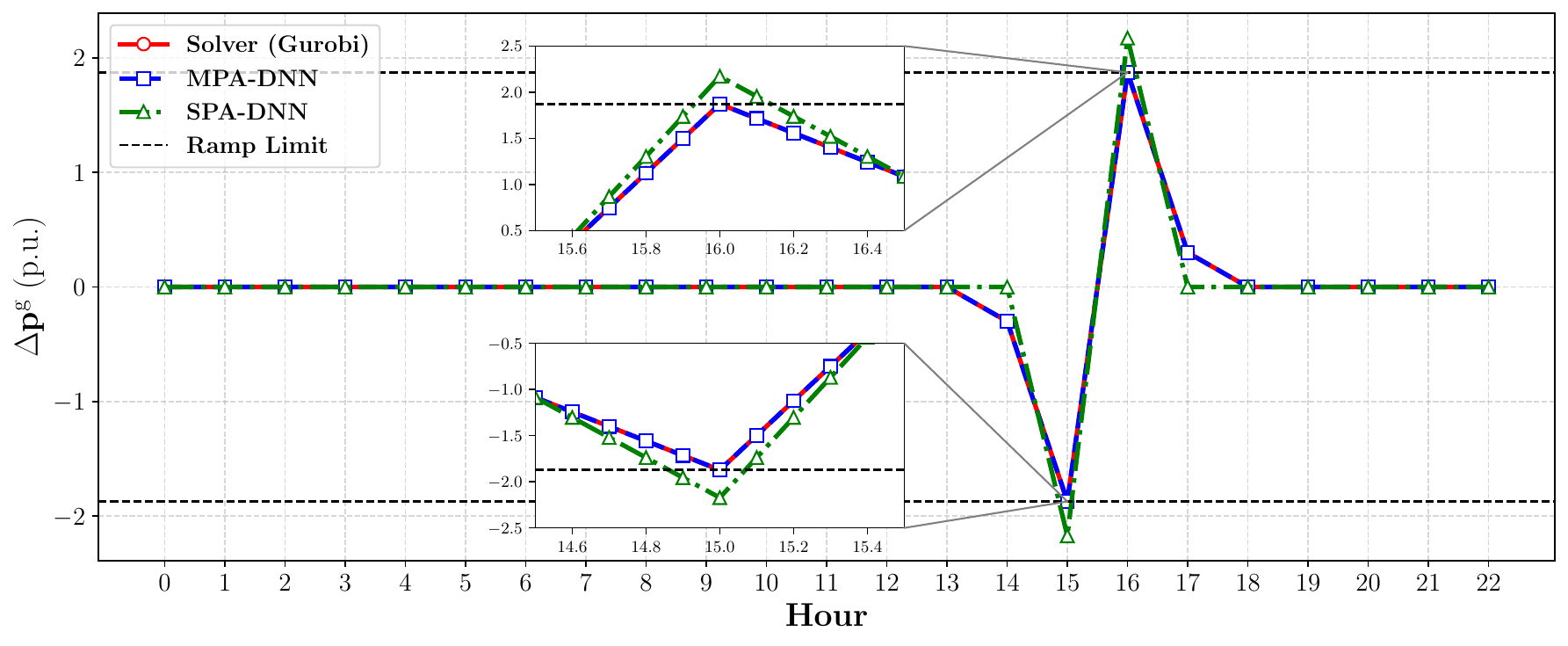}\vspace{-2mm}
    \caption{\small Ramping comparison for Generator 1 over a 24-hour horizon under load scale of 1.050. The figure shows that MPA-DNN and the solver produce feasible trajectories within the ramping limits, while SPA-DNN violates ramping constraints during hour 15–16.}
    \vspace{-5mm}
    \label{fig:fig2}
\end{figure}

\begin{table}[t]
    \centering
    \captionsetup{justification=centering, labelsep=period, font=footnotesize, textfont=sc}
    \caption{Generation costs at hours 15–17 under load scale of 1.050.}
    \label{table:hourly_costs_violation}
    \begin{tabular}{c|ccc}
        \toprule
        \textbf{Hour} & \textbf{Solver (Gurobi)} & \textbf{SPA-DNN} & \textbf{MPA-DNN (Proposed)} \\
        \midrule\midrule
        15 & 153,513.62 & 153,513.63 & 153,513.63 \\
        16 & 155,341.60 & \textbf{154,609.80} & 155,341.59 \\
        17 & 162,520.58 & 162,520.58 & 162,520.59 \\
        \bottomrule
    \end{tabular}
    \vspace{-5mm}
\end{table}

Fig.~\ref{fig:fig2} zooms in at hours 15-16, when the ramp rate constraint for Generator~1 becomes binding. As shown in the figure, both the solver and the proposed MPA-DNN determine the dispatch that is feasible at this hour. By contrast, since SPA-DNN determines the dispatch in each hour without considering ramping constraints, an infeasible schedule occurs during hours 15-16 at Generator~1.

Specifically, the dispatch from SPA-DNN at hour 16 achieves a slightly lower generation cost (\$154,610) than the corresponding DC-OPF solution (\$155,342), as summarized in Table~\ref{table:hourly_costs_violation}. However, this saving is achieved by ignoring the ramp limit, and is thus meaningless for real-world power-systems. These results imply the importance of modeling inter-temporal constraints in learning-based OPF methods, particularly when physically realizable solutions are required.

\section{Conclusion}
\color{black}In this paper, we proposed Multi-Period Projection-Aware Deep Neural Network (MPA-DNN), a novel deep learning framework for solving multi-period DC-OPF problems with inter-temporal constraints. \color{black} Unlike prior approaches that rely on supervised learning or ignore temporal feasibility (e.g., ramp limits), the proposed method embeds a differentiable projection layer into the model to explicitly modeling generator ramping limits and ESS operation and is trained via unsupervised learning framework. Experimental evaluations on the IEEE 39-bus system demonstrate that MPA-DNN achieves near-optimal performance while strictly satisfying physical constraints. Compared to supervised baselines, the proposed approach shows superior generalization under varying load conditions, maintaining both feasibility even under out-of-distribution scenarios. \color{black} For future work, we plan to extend the proposed method to more complex power system optimization problems, such as security-constrained AC-OPF and unit commitment. We also intend to provide a more detailed analysis of computational time, and to investigate a wider range of ESS configurations. \color{black}

\color{black}\section*{Appendix A}
$\mathbf{A} \in \mathbb{R}^{T \times p T}$, $\mathbf{b} \in \mathbb{R}^{T}$, $\mathbf{G} \in \mathbb{R}^{q \times p T}$, and $\mathbf{h} \in \mathbb{R}^{q}$ define the feasible region \(\mathcal{F}\). Here, $q = 6 p T + 2 p (T-1) + 2 n_\mathrm{l} T$ is the total number of inequality constraints. Here, $\otimes$ denotes the Kronecker product used to replicate base constraints across time or units. $\mathbf{D} \in \mathbb{R}^{(T-1) \times T}$ is a forward difference matrix encoding generator ramping behavior, and $\mathbf{S} \in \mathbb{R}^{T \times T}$ is a unit lower triangular matrix with ones. $\mathbf{p}^\mathrm{d} \in \mathbb{R}^{n_\mathrm{d} \times T}$ is the time-series active power demand matrix over all $T$ periods for $n_\mathrm{d}$ load buses, while $\mathbf{p}^{\text{ch}} \in \mathbb{R}^{n_\mathrm{e} \times T}$ and $\mathbf{p}^{\text{dis}}\in\mathbb{R}^{n_\mathrm{e} \times T}$ represent charging and discharging power of $n_\mathrm{e}$ ESS over $T$ periods. The block-selection matrices $\mathbf{U}_\mathrm{g} = [\mathbf{I}_{n_\mathrm{g}}, \mathbf{0}, \mathbf{0}]$, $\mathbf{U}_\text{ch} = [\mathbf{0}, \mathbf{I}_{n_\mathrm{e}}, \mathbf{0}]$, and $\mathbf{U}_\text{dis} = [\mathbf{0}, \mathbf{0}, \mathbf{I}_{n_\mathrm{e}}]$ extract generator outputs, ESS charging, and ESS discharging components, respectively, $\text{vec}(\mathbf{x})\in\mathbb{R}^{pT}$. Their time-replicated forms are designed as $\mathbf{V}_\mathrm{g} = \mathbf{I}_T \otimes \mathbf{U}_\mathrm{g}$, $\mathbf{V}_\text{ch} = \mathbf{I}_T \otimes \mathbf{U}_\text{ch}$, and $\mathbf{V}_\text{dis} = \mathbf{I}_T \otimes \mathbf{U}_\text{dis}$. We represent these constraints compactly as follows:\color{black}
\begin{subequations}
\begin{equation}
    \mathbf{A} = \mathbf{I}_T \otimes [\mathbf{1}_\mathrm{g}^\intercal, -\mathbf{1}_\text{ch}^\intercal, \mathbf{1}_\text{dis}^\intercal], 
    \mathbf{b} = \mathbf{1}_\mathrm{d}^\intercal \mathbf{p}^\mathrm{d}, \label{eq:mp_b}
\end{equation}
\begin{equation}\color{black}
    \mathbf{G} = 
    \begin{bmatrix}
    \mathbf{V}_\mathrm{g} \\
    - \mathbf{V}_\mathrm{g} \\
    \mathbf{V}_\text{ch} \\
    - \mathbf{V}_\text{ch} \\
    \mathbf{V}_\text{dis} \\
    -\mathbf{V}_\text{dis} \\
    \mathbf{D} \otimes \mathbf{U}_\mathrm{g} \\
    -\mathbf{D} \otimes \mathbf{U}_\mathrm{g} \\
    \mathbf{S} \otimes [\boldsymbol{\eta}^{\text{ch}} \mathbf{U}_\text{ch} - \mathbf{U}_\text{dis} / \boldsymbol{\eta}^{\text{dis}}] \\
    - \mathbf{S} \otimes [\boldsymbol{\eta}^{\text{ch}} \mathbf{U}_\text{ch} - \mathbf{U}_\text{dis} / \boldsymbol{\eta}^{\text{dis}}] \\
    \mathbf{I}_T \otimes [\mathbf{\Phi} \mathbf{M}^\mathrm{g}, - \mathbf{\Phi} \mathbf{M}^\text{e}, \mathbf{\Phi} \mathbf{M}^\text{e}] \\
    - \mathbf{I}_T \otimes [\mathbf{\Phi} \mathbf{M}^\mathrm{g}, - \mathbf{\Phi} \mathbf{M}^\text{e}, \mathbf{\Phi} \mathbf{M}^\text{e}]
    \end{bmatrix}, \label{eq:mp_G}
\end{equation}
\begin{equation}\color{black}
    \mathbf{h} = 
    \begin{bmatrix}
    \text{vec}(\mathbf{1}_T \otimes [\overline{\mathbf{p}}^{\mathrm{g}}, \mathbf{0}, \mathbf{0}]^\intercal) \\
    \text{vec}(-\mathbf{1}_T \otimes [\underline{\mathbf{p}}^{\mathrm{g}}, \mathbf{0}, \mathbf{0}]^\intercal) \\
    \text{vec}(\mathbf{1}_T \otimes [\mathbf{0}, \overline{\mathbf{p}}^{\text{ch}}, \mathbf{0}]^\intercal) \\
    \text{vec}(\mathbf{1}_T \otimes [\mathbf{0}, \mathbf{0}, \mathbf{0}]^\intercal) \\
    \text{vec}(\mathbf{1}_T \otimes [\mathbf{0}, \overline{\mathbf{p}}^{\text{dis}}, \mathbf{0}]^\intercal) \\
    \text{vec}(\mathbf{1}_T \otimes [\mathbf{0}, \mathbf{0}, \mathbf{0}]^\intercal) \\
    \text{vec}(\mathbf{1}_{T-1} \otimes [\mathbf{R}^{\text{u}}, \mathbf{0}, \mathbf{0}]^\intercal) \\
    \text{vec}(\mathbf{1}_{T-1} \otimes [\mathbf{R}^{\text{d}}, \mathbf{0}, \mathbf{0}]^\intercal) \\
    \mathbf{1}_T \otimes \overline{\mathbf{E}} \\
    \mathbf{1}_T \otimes \underline{\mathbf{E}} \\
    {\color{black}{\text{vec}\Big(\mathbf{\Phi} \big(\mathbf{M}^\mathrm{d} \mathbf{p}^\mathrm{d}\big)\Big)}} + \mathbf{1}_T \otimes \overline{\mathbf{p}}^\mathrm{l} \\
    {\color{black}{ - \text{vec}\Big(\mathbf{\Phi} \big(\mathbf{M}^\mathrm{d} \mathbf{p}^\mathrm{d}\big)\Big)}} + \mathbf{1}_T \otimes \overline{\mathbf{p}}^\mathrm{l}
    \end{bmatrix}.
\end{equation}
\end{subequations}
\vspace{-2mm}
\bibliographystyle{IEEEtran}
\bibliography{reference.bib}
\end{document}